# Anomalous Absorption in Arrays of Metallic Nanoparticles: A Powerful Tool for Quantum Dot Optoelectronics

*Augustin Caillas[1], Stéphan Suffit[1], Pascal Filloux[1], Emmanuel Lhuillier[2], Aloyse Degiron[1,*]*

[1]Université de Paris, CNRS, Laboratoire Matériaux et Phénomènes Quantiques, F-75013 Paris, France

[2]Sorbonne Université, CNRS, Institut des NanoSciences de Paris, INSP, F-75005 Paris, France

ABSTRACT: Periodic arrays of noble metal nanoparticles are emblematic nanostructures in photonics. Their ability to sustain localized surface plasmon resonances has been used throughout the years to demonstrate a variety of passive and active functionalities such as enhanced luminescence in dipolar media and LEDs as well as higher responsivities in photoconductive detectors. Here, we show that additional magnetic resonances, associated with inductive current loops between the nanoparticles and accessible with transverse electric waves, emerge in the limit of dense arrays with subwavelength periods. Moreover, their interplay with the plasmons of the system results in spectrally sharp analogs of electromagnetically induced absorption (EIA). We use these metasurfaces to induce changes and enhancements in the emission, absorption, photoconduction and polarization properties of active layers of PbS nanocrystals, illustrating the potential of EIA beyond the passive functionalities demonstrated so far in the literature.

MAIN TEXT

In most instances, the optical properties of periodic arrays of noble metal nanoparticles are governed by the plasmonic resonances sustained by each metallic inclusion. The main difference between these resonances and those sustained by isolated nanoparticles is that they are broadened and shifted to longer or smaller wavelengths due to the weak coupling between adjacent unit cells[1]. At resonance, the scattering cross-section of the nanoparticles is maximized, resulting in transmission dips that make the structures usable as passive optical absorbers or filters[2,3]. Moreover, the high local fields associated with plasmonic arrays have been used to produce strong surface enhanced Raman scattering[4], increase the luminescence output of dipolar media[5,6] and endow quantum dot LEDs with advanced functionalities[7]. Conversely, these structures can also enhance the absorption cross-section within solar cells[8] and infrared detectors[9–12], considerably improving the responsivities of the devices.

Interestingly, the properties of periodic plasmonic arrays are known to change dramatically in the special case of nanoparticles plunged in an (almost) symmetric environment and at wavelengths allowing the emergence of non-zeroth diffraction orders. Under these very specific conditions, the coupling to grazing diffraction orders and the plasmonic resonances lead to the formation of lattice surface modes[13–16]. These dispersive and polarized modes are so narrow (half width at half maximum of a few nanometers) and low-loss that they can dramatically enhance and polarize the luminescence of dipolar media[17] and even lead to lasing[18] and photonic Bose-Einstein condensation in such active layers[19]. Here, we show that there exists another interesting special case that emerges when the particle inter-distance is very small, regardless of whether the local environment surrounding the plasmonic nanoparticles is symmetric or not.

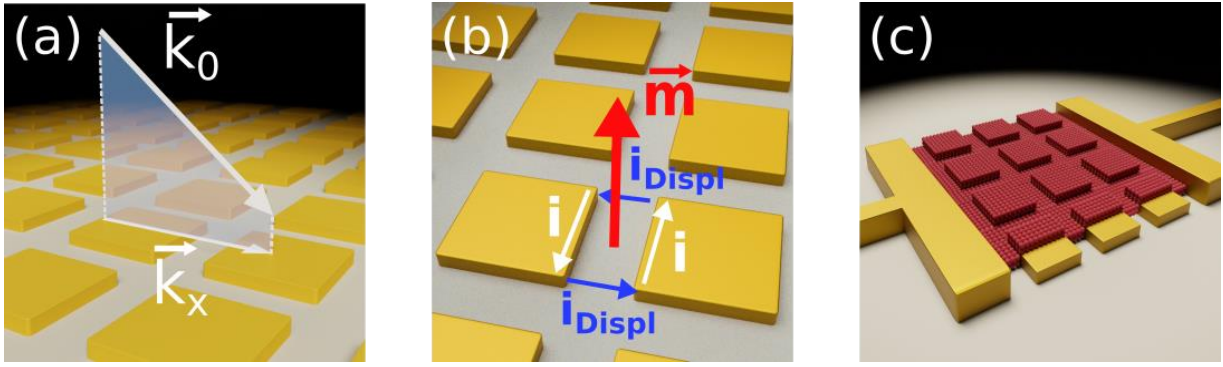

**Figure 1.** (a) Schematic of the simulated periodic array of square Au nanoparticles. The white arrows represent the wavevector of the incident light ($\vec{k}_0$) and its projection along the x-axis ($\vec{k}_x$). (b) Illustration of the currents *i* and displacement currents $i_{Displ}$ that form a resonant loop associated with a magnetic moment **m** normal to the surface. Only one current loop is shown for clarity. This pattern repeats itself between all adjacent metallic edges. (c) Simplified view of the photoconductive detector studied on Figure 3. The actual device is made of interdigitated electrodes defining 10 µm wide channels filled with square Au nanoparticles and covered with 2-3 monolayers of PbS NCs (represented as red spheres).

We start this study with finite element simulations of the periodic array of Au nanoparticles depicted on Figure 1a. The structure is supported by a dielectric substrate with refractive index $n_g = 1.45$ (corresponding to the value of the glass substrates used in our experiments), the nanoparticles are 25 nm thick, they have a square shape of 400 nm by 400 nm and the period is P = 600 nm. The spatial extension of the array is taken as infinite so as to reduce the computational domain to a single periodic unit cell (see simulation details in Section I of the Supporting Information). Figure 2a shows the simulated reflectance spectrum when the structure is illuminated by a plane wave under normal incidence from the air side (dark blue curve). A single peak, corresponding to the excitation of localized plasmon resonances that are considerably broadened due to the closely spaced square nanoparticles[1], is observed. If one repeats the simulations at an incidence angle of 20°, a more complex spectrum is observed for the TE polarization, with a pronounced dip encroaching on the plasmon resonance at 1280 nm (red curve). The dip occurs for all non-zero incidence angles and is largely dispersionless, as can be appreciated on the dispersion map of Figure 2c where the reflection for TE-polarized waves is plotted as a function of both the photon energy $\hbar\omega$ and the in-plane component $k_x$ of the incident wavevector. Furthermore, its magnitude evolves from a small dent in the reflectance peak at approximately $\hbar\omega = 0.97$ eV (1280 nm) to a pronounced minimum that shifts to smaller energies (longer wavelengths) as the incidence angle increases.

To gain further insight, we plot the absorption spectra and dispersion relation for the same cases in Figures 2b,2d. A sharp peak, located at the same wavelength as the dip in reflection, emerges on top of the broad plasmonic resonances for all non-zero angles.

These results are unexpected for a simple array of Au nanoparticles because the latter is supposed to exhibit a single reflection maximum caused by the weakly-coupled plasmonic resonances[1]. The formation of a narrow reflection dip and an equally sharp absorption peak cannot be attributed to diffraction effects because the onset of non-zeroth diffraction orders, defined by $\hbar\omega = \pm\hbar k_x c/n_g \mp \hbar 2\pi c/(n_g P)$ (dashed white lines on Figure 2c), is above the frequency range where it occurs (except at large incidence angles where diffraction renders the dip slightly dispersive). In other words, we operate the nanoparticle array in the subwavelength regime, i.e., as a metasurface.

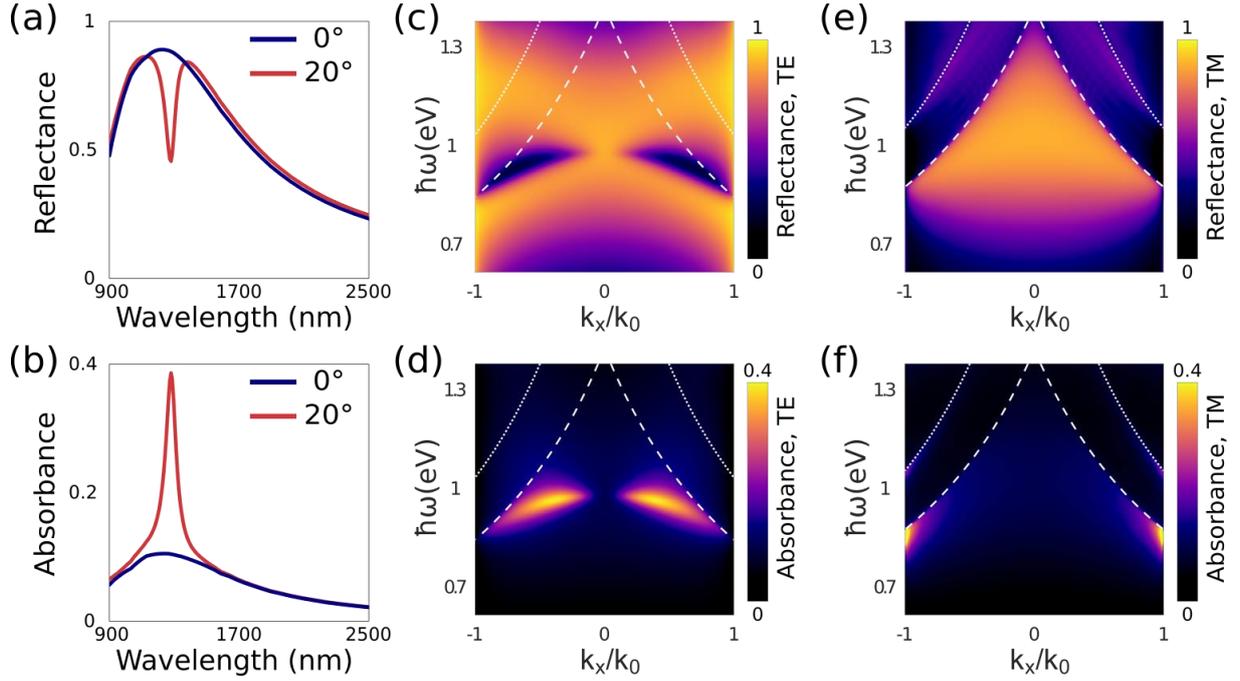

**Figure 2.** Simulated optical properties of the periodic array of square Au nanoparticles. (a) TE-polarized reflectance spectra at normal incidence and for an oblique incidence of 20° degrees. (b) Corresponding absorbance spectra. (c) and (d) TE-polarized reflectance and absorbance spectra plotted as a function of $k_x/k_0$ and the photon energy $\hbar\omega$. The white dashed lines are the light lines in the substrate ($n_g=1.45$) and the white dotted lines are the light lines in air. (e) and (f) TM-polarized reflectance and absorbance spectra plotted as a function of $k_x/k_0$ and the photon energy $\hbar\omega$.

These unusual features disappear when the filling factor of the structure decreases, as shown on Figure S2 for an array with the same period of 600 nm but made of Au nanoparticles with a reduced side of 200 nm. In other words, the reflection dip and absorption peak evidenced on Figures 2a-2d are related to the close proximity between the edges of adjacent metallic patches. This conclusion is supported by an examination of the fields and current density lines around the structure (Supporting Information, Figures S3-S5). The latter reveals that in addition to the plasmonic resonances sustained by the nanoparticles, an array of magnetic LC resonances, characterized by inductive current loops formed by currents $i$ within the metal and pure displacement currents $i_{Disp}$ between the metallic patches, is excited for off-normal incidence angles. Such magnetic resonances occurring in the near-infrared and visible range are well known to the metamaterial community where they play a central role in the realization of a negative effective permeability in coupled nanorods[20] and fishnet metamaterials[21]. Their resonance frequency can be tuned with the inductance (and thus the length of the current loops, defined by the lateral size of the patches and the opening between them) and the capacitance (essentially defined by the opening between the patch). Their magnetic dipole moment $\vec{m}$, oriented perpendicular to the current loops, is normal to the plane of the structure in our case, as sketched on Figure 1b.

We can now understand the results of Figures 2a-2d. The fact that the excitation of the LC resonances creates a sharp dip in the reflectance peak of the surface plasmons indicates that the two sets of modes interfere destructively, reducing the scattering of the structure while increasing its absorbance. In the special case of normal-incidence illumination, the LC resonances cannot be excited because the incident magnetic field is perpendicular to their moment $\vec{m}$. Thus, the spectra computed at 0° in Figures 2a-2b only feature the signature of the surface plasmons. Because of the same selectivity rule, the LC resonances do not occur for the TM polarization, a point that can be verified by examining the reflectance and absorption spectra computed for this polarization on Figures 2e-2f (the features seen in reflectance closely follow the light lines and are therefore diffractive effects that have little influence on the absorption spectra except at grazing incidence).

Similar resonances producing sharp absorption peaks have been reported in the past by authors[22–26] who sought to demonstrate photonic analogs of electromagnetically induced absorption (EIA)—a narrow absorption resonance of certain atomic systems coherently interacting with pump-probe fields[27]. A key ingredient for atomistic EIA is either a transfer of coherence from two energy levels or a transfer of population from the ground level to a reservoir[28]. In photonic analogs of EIA, this transfer is played by the interferences between optical modes having different linewidths. More precisely, the few demonstrations of photonic

EIA that have been made across the electromagnetic spectrum have relied on pairing one type of optical resonator with another cavity or waveguide that does not directly interact with the incoming radiation[22–26]. Here, in contrast, we work with a single class of metallic nanopatches and a photonic analog of EIA appears as a consequence of their mutual coupling in the limit of highly compact structures (see Supporting Information, Section IV and Figure S6 for additional considerations on the design rules and how to obtain the same behavior at other wavelengths). Moreover, the two resonances involved in the process are both efficiently excited by the incoming radiation.

We now confirm these numerical results with experiments that also demonstrate the potential of such metasurfaces for optoelectronics applications. The only differences with the structure studied on Figure 2 is that the square array of Au nanoparticles occupies a finite footprint of 200 x 200 $\mu m^2$ and that it is covered by a continuous layer of PbS nanocrystals (NCs) cross-linked with ethanedithiol (EDT, see Supporting Information). This thin (15-20 nm) layer does not fundamentally change the properties of the nanoparticle array: the reflectance spectrum of the sample, plotted on Figure 3a (and Figure S9 over a wider spectral range), exhibits a dip in the middle of the plasmonic resonance, just as the simulated structure without PbS NCs of Figure 2. These spectral features occur at slightly longer wavelengths because the thin layer of PbS NCs effectively increases the refractive index around the metallic squares. Moreover, the dip is less pronounced because the idealized plane wave illumination of Figure 1 cannot be implemented in practice. In our experiments, the sample is illuminated with unpolarized white light under a range of incidence angles between 12° and 23.6° using a 15X reverse Cassegrain objective and the resulting reflectance spectrum agrees well with the model when the effect of the PbS NCs and the experimental illumination conditions are taken into account (compare the continuous and dashed purple curves).

We turn to the photoluminescence (PL) properties of the sample (Figure 3b). As shown in a previous article[29], the optoelectronic properties of PbS NCs cross-linked with EDT are governed by their thermalized excitons, implying that the PL intensity $I(\lambda)$ and the resonant excitonic absorption cross-section $A_{RES}(\lambda)$ are related by a local Kirchhoff law of the form $I(\lambda) \propto A_{RES}(\lambda) \cdot W(\lambda,T)$, where $W(\lambda,T) = 2hc^2\lambda^{-5}\exp[-hc/(\lambda k_B T)]$ is the Wien approximation of Planck's law, h and $k_B$ are the Planck and Boltzmann constants, $\lambda$ is the wavelength and T is the carrier temperature.

Without the metasurface, the PL spectrum of the NC layer on glass features an inhomogeneously-broadened peak in the near-infrared (note that the exact shape and position of the peak depends on the excitation power because the latter determines the carrier temperature, and therefore the Wien term $W(\lambda,T)$, in the above equation[29]). With the metasurface, a general PL enhancement is observed and a sharp peak appears at the same wavelength as the EIA dip of Figure 2a. This result suggests that the optical analog of EIA has strongly increased the absorption cross-section $A_{RES}(\lambda)$ of the NCs covering the metasurface, which in turns has led to a sharp enhancement of the PL at the same wavelengths.

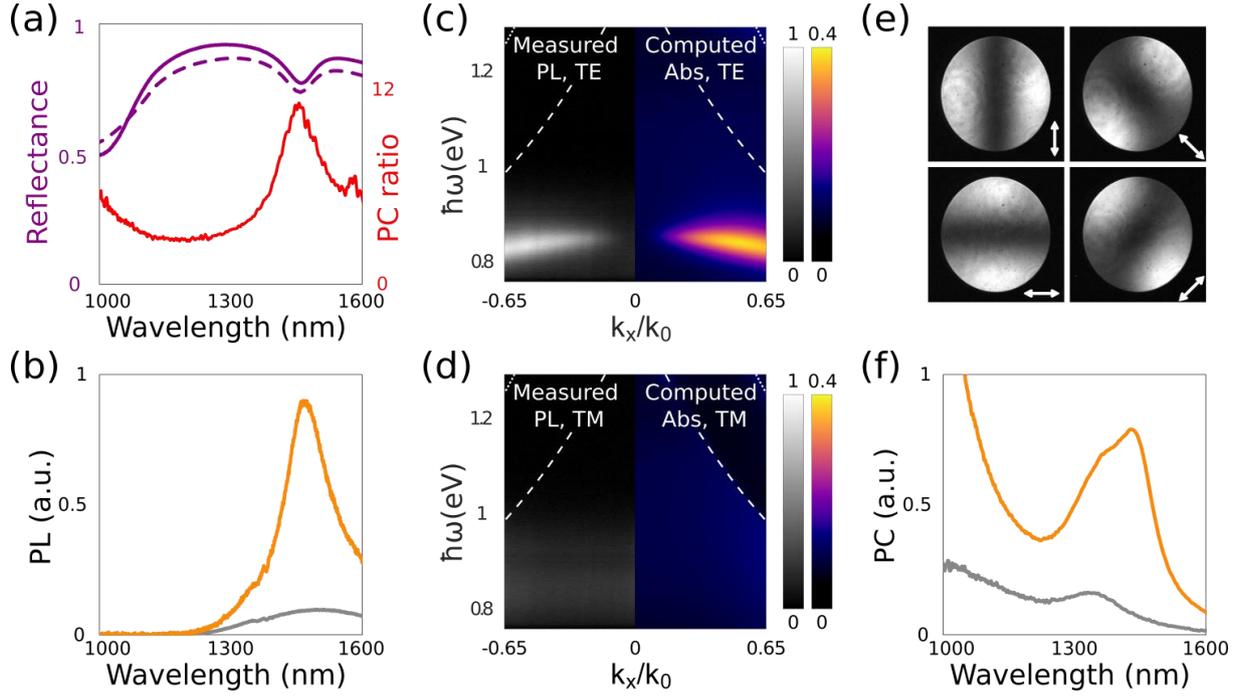

**Figure 3.** (a) Experimental and simulated reflectance spectra of the sample (continuous and dashed purple curves, respectively). (b) PL spectrum of the PbS NCs above and outside the nanoparticle array (yellow and gray curve, respectively). (c) Experimental TE-polarized PL dispersion relation (grayscale colormap in arbitrary units) vs. simulated TE-polarized absorbance dispersion relation (dark blue to yellow colormap). (d) Same as panel (c), but for the TM polarization. The colormap scales are the same as for the TE-polarization. (e) Back focal plane images of the PL at 1450 nm (plotted in the $k_x/k_0, k_y/k_0$ space), taken after inserting a linear polarizer in the beam path. The orientation of its polarization axis is indicated by the double arrows. The outer diameter of the images has a value of 0.65 corresponding to the numerical aperture of our 50X objective. (f) Photocurrent spectra of PbS NC photoconductive detectors with (yellow curve) and without (gray curve) Au nanoparticles patterned between the interdigitated electrodes. The ratio between these two spectra is the red curve of panel (a).

To confirm this hypothesis, we have measured the dispersion of the emitted PL as a function of $\hbar\omega$ and $k_x$ and for two orthogonal polarizations (all the relevant experimental details are given in section V of the Supporting Information). When the polarization is perpendicular to $k_x$, corresponding to TE polarized waves in this specific plane of emission, the sharp emission peak is observed and it closely follows the dispersion of the EIA dip in absorption predicted by the model (Figure 3c). For the orthogonal TM polarization, this peak does not appear in the emission spectrum, which is again consistent with the simulated absorption (Figure 3d). These findings do not only provide an experimental confirmation of EIA, but also illustrate its potential for influencing the absorption and luminescence properties of an active dipolar medium. In comparison, the plasmonic resonances alone (i.e. outside the EIA window for the TE polarization and at all angles for the TM polarization) have a weak effect.

Due to the symmetry of the structure, the experimental results plotted on Figures 3c-3d remain true for a 90-degree rotation of the in-plane wavevector, while all other angles are a linear superposition of these two cases. Figure S10 confirms this point by imaging the in-plane wavevector distribution of the PL emitted at 1450 nm (note that Figures 3c, 3d and S10 are nothing else than different planes of the $k_x/k_0, k_y/k_0, \omega$ space). A broad cone is observed, illustrating the emission diagram of the structure caused by EIA at this wavelength. Moreover, since EIA only occurs for TE-polarized waves, the emission beam has an azimuthal electric field distribution, as evidenced on Figure 3e where the same in-plane wavevector distribution of the emitted PL at 1450 nm is imaged through a linear polarizer rotated between 0 and 180 degrees. Such non-trivial polarization features are worth noticing considering that we do not work with laser beams, which can be manipulated at will with optical elements that take advantage of their spatial coherence, but with the spontaneous emission of an assembly of incoherent quantum emitters. Importantly, the PL properties do not depend on the size of the excitation spot: while all the data represented on Figure 3 have been obtained by pumping a tiny fraction of the structure with a diffraction-limited HeNe laser spot, Figure S11 demonstrates that the same results hold true if the $200 \times 200$ µm$^2$ of the sample are illuminated almost in full with visible white light.

Finally, we demonstrate the potential of the metasurface for enhanced photodetection. Since the optoelectronic properties of our layers of PbS NCs are governed by a local Kirchhoff law (see above), the sharp PL peak that occurs at the EIA condition is equivalent to a strong enhancement of their excitonic absorption cross-section $A_{RES}$ *at the same wavelength*. In other words, by illuminating the sample with oblique TE-polarized waves, the electromagnetic energy is not wasted as heat in the metal but transferred to the NCs—which in turn should enhance their photoconductive properties. To verify this point, we have fabricated a new sample in which the square nanoparticles are now patterned between planar interdigitated Au electrodes. The geometric parameters of the metasurface are the same as before and the sample is also covered with a 15-20 nm thick layer of PbS NCs cross-linked with EDT (see schematic of Figure 1c). Figure 3f shows the spectrum of the photocurrent flowing between the electrodes when the sample is illuminated with an external white light source. The data have been collected with a Fourier transform photocurrent spectroscopy setup described in the Supporting Information, using the same reverse Cassegrain objective as the one used for measuring the reflectance spectrum of Figure 2a. The photocurrent (PC) (yellow curve) exhibits a general enhancement and an additional narrow peak at 1460 nm compared to a control sample without Au nanoparticles (gray curve). We attribute the overall enhancement to an improved bottom reflectivity induced by the metasurface, allowing part of the incident light that has passed through the thin NC layer without being absorbed to contribute to the photocurrent after being backscattered by the Au nanoparticles. As for the new peak, it can be unequivocally attributed to EIA due to its wavelength position and narrow width. To confirm this point, we plot on Figure 3a the ratio between the PC spectra measured with and without metallic nanoparticles. The curve exhibits a sharp peak at the EIA condition where the enhancement exceeds 10. Moreover, the peak coincides almost perfectly with the position of the sharp PL maximum of Figure 3b, further demonstrating the equivalence between absorption and emission in our layers of thermalized NCs. In the Supporting Information, we provide another experimental example in which the NC absorption is enhanced at smaller wavelengths using a metasurface with reduced feature sizes (Figure S12).

In conclusion, a structure as simple as an array of square Au nanoparticles can behave as a metasurface with a surprisingly rich and technology-enabling behavior when properly designed at a subwavelength scale. All the effects evidenced in this study are a consequence of the optical analog of EIA that develops within the Au nanoparticle array. Different from earlier demonstrations in the literature, which focused on more complex structures used as passive filters and absorbers that turned photonic analogs of EIA into Joule losses[22–26], we have shown that the absorption can be leveraged to induce dramatic changes in the optoelectronic properties of PbS NC assemblies. As a corollary, the strong enhancements and sharp spectral modifications of the emission, absorption and photocurrent spectra, the demonstration of a directional and azimuthally polarized emission, as well as the direct relationship of all these observations with EIA, demonstrate that the relevant figure of merit to tune the optoelectronic properties of an active medium with thermalized carriers is its absorption cross-section—and not the local density of photonic states as is usually the case for single emitters interacting with nanostructures. In this respect, these results are not restricted to ensemble of PbS NCs and should hold for any active medium with thermalized carriers, including quantum wells, semiconductors, two-dimensional materials and organic layers.

ASSOCIATED CONTENT

**Supporting Information**.

The following files are available free of charge.

Details on the numerical simulations; Schematic cross-section of the model used to simulate a Au particle array covered with a thin layer of PbS NCs (Figure S1); Dispersion relation of an array with smaller square Au patches (Figure S2); Analysis of the fields supported by the metasurface when EIA occurs (Figure S3); Charge distributions at the EIA wavelength (Figure S4); Current density lines evaluated at the EIA condition and outside the EIA condition (Figure S5); How to design the metasurfaces to operate at other wavelengths (Figure S6); Experimental details (chemicals, PbS NC synthesis, sample fabrication, PL experiments and PC experiments); Scanning electron micrograph of an actual sample (Figure S7); Schematic of the PL experimental setup (Figure S8); Experimental and simulated reflectance spectrum of the sample studied on Figure 2a, plotted over an extended wavelength range (Figure S9); Back focal plane at 1450 nm without polarizer inserted in the beam path (Figure S10); PL experiments realized with a wide field pumping (Figure S11); Simulated and experimental data (reflectance, photocurrent spectrum) for an array with smaller feature sizes (Figure S12).

ACKNOWLEDGMENTS

We acknowledge support from the European Research Council grant FORWARD (reference: 771688). This work was supported by French state funds managed by the ANR through grant Copin (ANR-19-CE24-0022).

CORRESPONDING AUTHOR
* aloyse.degiron@u-paris.fr

# Anomalous Absorption in Arrays of Metallic Nanoparticles: A Powerful Tool for Quantum Dot Optoelectronics

Augustin Caillas[1], Stéphan Suffit[1], Pascal Filloux[1], Emmanuel Lhuillier[2], Aloyse Degiron[1,*]

[1]Université de Paris, CNRS, Laboratoire Matériaux et Phénomènes Quantiques, F-75013 Paris, France

[2]Sorbonne Université, CNRS, Institut des NanoSciences de Paris, INSP, F-75005 Paris, France

*aloyse.degiron@u-paris.fr

# Supporting Information

# I. Numerical model

The structures are simulated using the RF module of Comsol Multiphysics, a finite element solver. The model studied in Figure 1 consists of a single unit cell flanked with periodic conditions. The bottom half of the unit cell has a height of 3 µm and is filled with glass with a refractive index $n_g = 1.45$. This index corresponds to the experimental value of our Borofloat glass substrate from Plan Optik, measured by fitting the diffraction orders observed in some of our PL maps with the light lines $\omega = \pm k_x c/n_g \mp 2\pi c/(n_g P)$. The upper half is also 3-µm high and is filled with air (n = 1). At the interface between these two media, the Au antennas are modelled with a permittivity taken from the experimental literature.[1] Electromagnetic waves enter and exit the model through ports applied on the top and bottom of the computational domain. The number of ports on each side is chosen so as to take all diffraction orders into account. The port associated with the zeroth diffraction order on the air side is used to launch a TE- or TM-polarized wave at a specified incidence angle.

The simulations presented in Figure 2a utilize a slightly modified version of the model to account for the actual dimensions of the Au squares (415 nm rather than 400 nm, see Figure S7) and for the presence of the PbS nanocrystals (NCs) covering the sample. By systematic comparisons between experiments and simulations, we have noticed that the influence of the PbS NCs on the optical properties of the arrays can be modeled in a satisfying manner by burying the square nanoparticles 10 nm below the surface of the glass substrate (the refractive index is still $n_g = 1.45$ without imaginary part). We have verified that this approximation does not only work for the specific structure studied in the article, but also for other metasurfaces operating at other wavelengths, as can be appreciated for example on Figure S12.

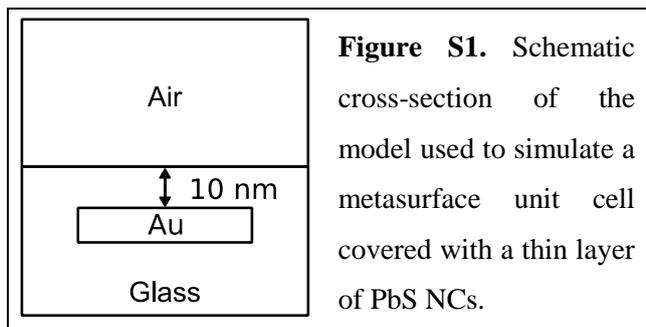

**Figure S1.** Schematic cross-section of the model used to simulate a metasurface unit cell covered with a thin layer of PbS NCs.

## II. Numerical simulations with a much smaller filling factor

In these simulations, the period P= 600 nm is the same as in the main text but the lateral size of the Au particles has been reduced to 200 nm × 200 nm (the Au thickness has been kept constant at 25 nm). All other conditions are the same as in Figure 2. Contrarily to the EIA case studied in the main text, there is no dip in the reflectance peak for the TE polarization. Moreover, the absorption maxima are essentially similar to those seen on the reflectance map, which is the standard and expected behavior of a nanoparticle array with optical properties dominated by plasmonic resonances localized on each Au inclusion. Note also that these plasmonic resonances are shifted to higher frequencies due to the smaller size of the Au nanoparticles and that they are strongly modulated by the diffraction orders that arise above the light lines.

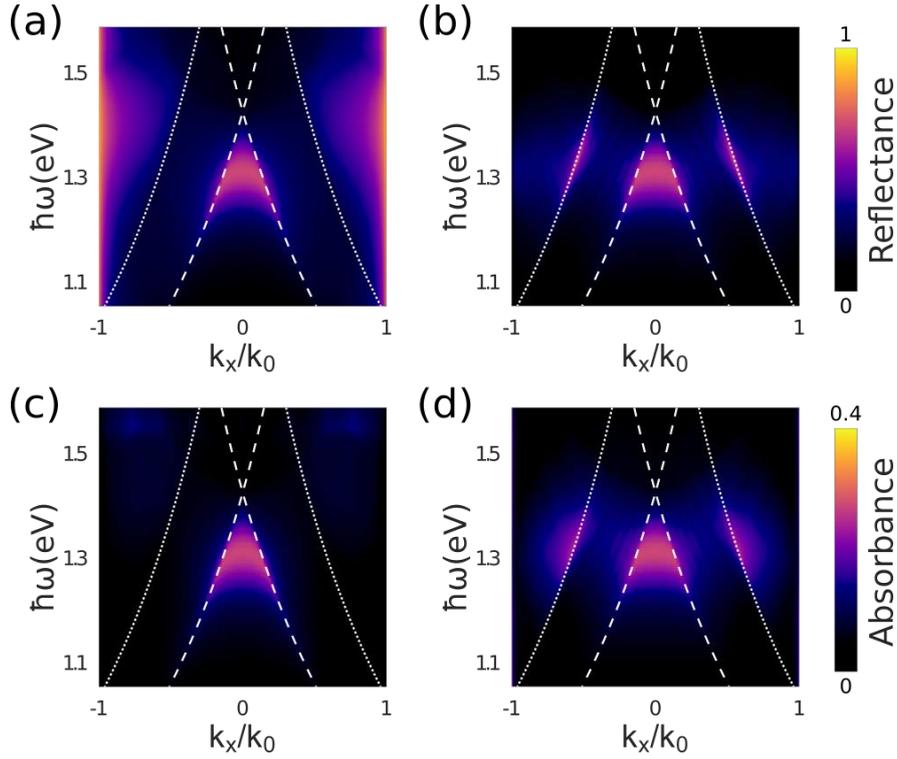

**Figure S2.** Simulated optical properties of the array when the size of the square nanoparticles is reduced to 200 nm × 200 nm × 25 nm. (a)-(b) Simulated reflectance dispersion relation for the TE and TM polarizations. (c)-(d) Simulated absorbance for the TE and TM polarizations. The dashed and dotted white lines are the light lines in the glass substrate and air, respectively.

## III. Analysis of the fields and associated currents when EIA occurs

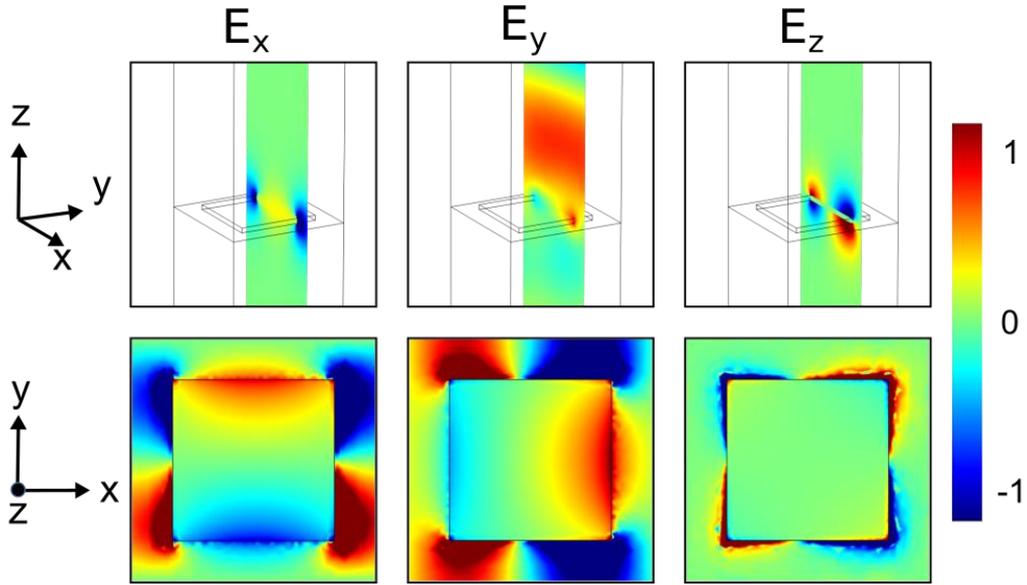

**Figure S3.** Plots of the $E_x$, $E_y$ and $E_z$ electric field components in the x-z plane (top row) and x-y plane (bottom row) at the wavelength of the EIA resonance ($\lambda$=1280 nm). The colors have been saturated to improve the visibility of the field symmetries. The structure shown here is the same as in Figure 2.

The spatial variations of the three components of the instantaneous electric field at the EIA wavelength are plotted in Figure S3 in two orthogonal planes. The positive (resp. negative) values mean that these *x*, *y* or *z* components are oriented along the positive (resp. negative) *x*-, *y*- or *z*-axes. From these maps, it is possible to deduce a snapshot of the charge distribution, based on the consideration that the electric field lines, which are tangential to the electric field vector at all points, always originate from positive charges and terminate at negative charges:

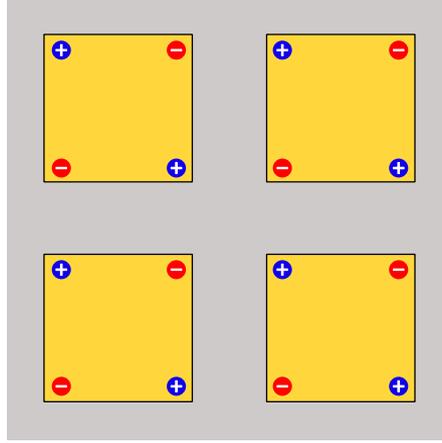

**Figure S4.** Schematic of the charges in the *x-y* plane.

The associated current distribution are the loops sketched in the main text on Figure 1b. Note that these current loops are formed by real currents flowing on the edges of the Au patches and displacement currents $\vec{\imath}_{Disp} \propto \partial \vec{D}/\partial t$ within and between the Au patches, as is well documented in the literature for related geometries such as pairs of metallic rods and fishnet metamaterials.[2,3]

To gain quantitative insight, we plot the actual current density lines predicted by the numerical model on Figure S5. Notice that current loops are only observed when the EIA peak occurs (case B in Figure S5). These loops are resonant because they disappear outside the EIA peak (cases A and C). Each loop has a mirror symmetry with respect to the horizontal *x*-axis. The loops do not possess a mirror symmetry in the other direction because the plane-of-incidence for oblique illumination is the *x-z* plane (see notations on Figure 1 in the main text).

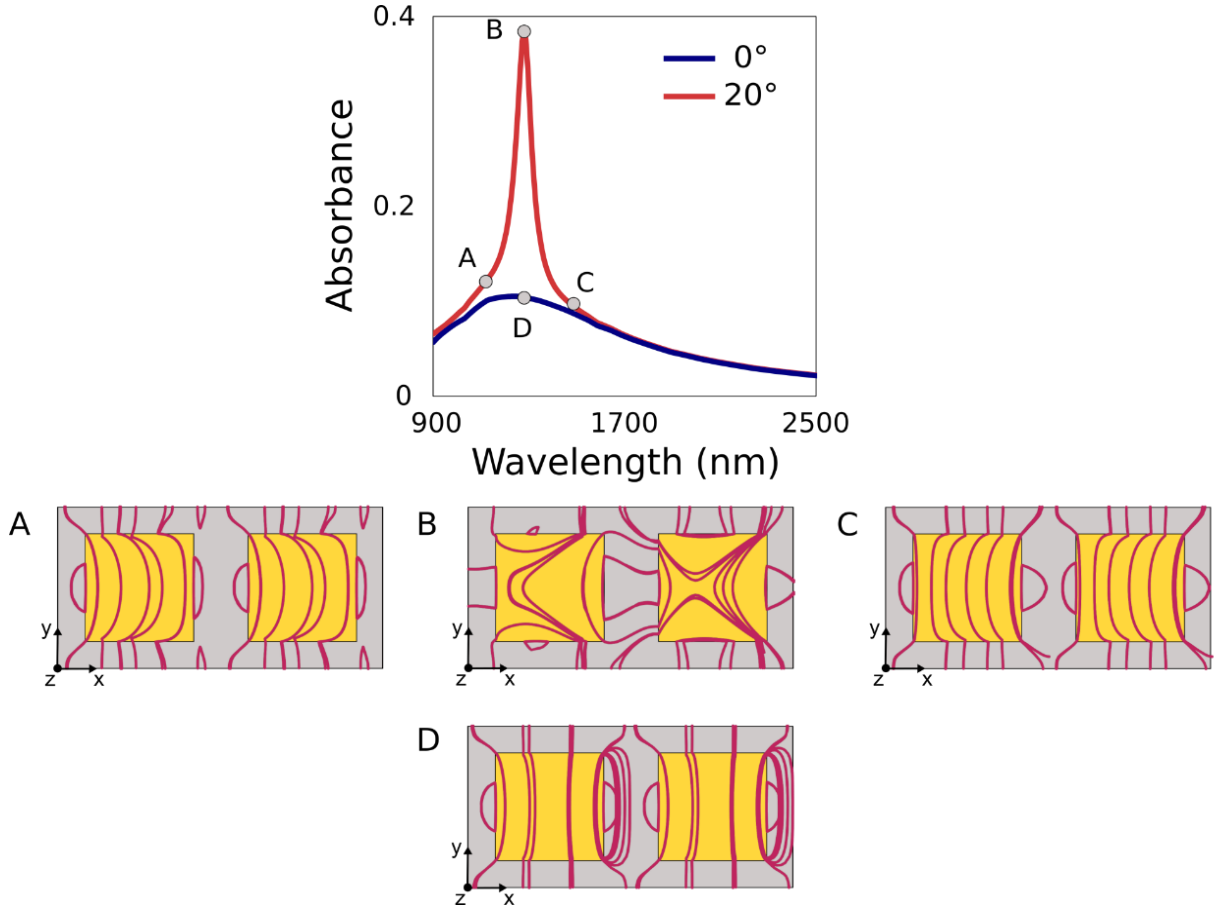

**Figure S5.** Current density lines computed for different wavelengths (A: 1120 nm, B and D: 1280 nm, C: 1490 nm) and different incidence angles (A, B & C: illumination at 20° with TE-polarized plane waves, D: illumination at normal incidence). The structure is the array of Au patches examined in Figure 2 in the main text. The absorbance spectra of Figure 2b are reproduced here with gray marks identifying the four cases A, B, C and D.

## IV. How to design the metasurfaces to operate at other wavelengths

To produce a Fano response analogous to EIA, the plasmonic and LC resonances must overlap. The main geometrical parameter affecting the spectral position of the plasmonic resonance is the side of the Au patches because these plasmons are standing waves across the patches. However, this geometrical parameter also affects the size of the current loops between adjacent Au patches and thus the inductance of the LC resonances. In contrast, the period P of the system has a relatively small influence on the position of the plasmon resonances (their spectral width, however, is very sensitive to this parameter[4]) but it has a stronger impact on the capacitance of the LC resonance. In the case of an ideal lossless and perfectly conducting LC resonator, its resonance wavelength is proportional to $\sqrt{LC}$. This result, however, cannot be directly applied to our system due to the strong retardation effects occurring in metals at optical frequencies as well as the complex inhomogeneous environment in which the LC resonances develop. We have heuristically found that an efficient way to shift the EIA peak to smaller (resp. larger) wavelengths is to reduce (resp. increase) the side of the Au patches and the value of the period simultaneously, but not necessarily with a strict scaling factor, as illustrated on Figure S6.

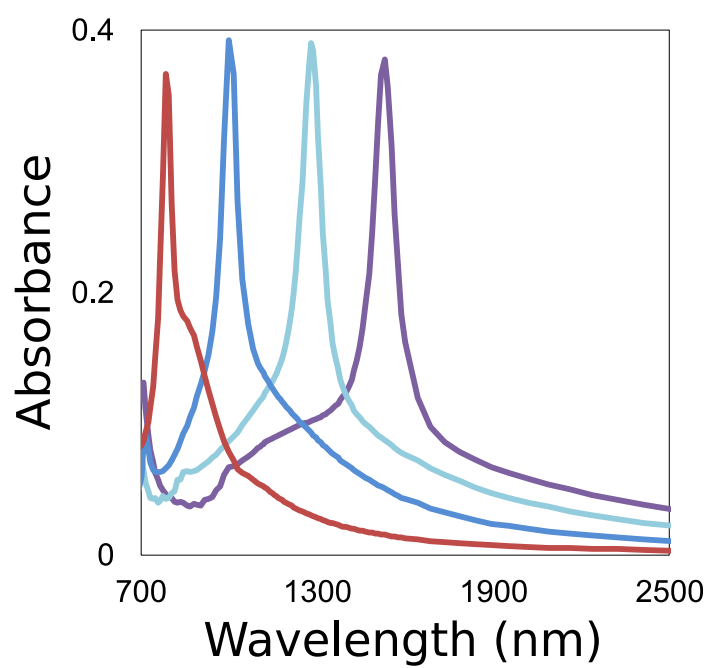

**Figure S6.** Numerical simulations of four different arrays of Au patches with different sides S and periods P. Red curve: P = 400 nm and S = 200 nm. Blue curve: P = 500 nm and S = 300 nm. Light blue curve: P = 600 nm and S = 400 nm. Violet curve: P = 700 nm and S = 500 nm. The metal thickness is 25 nm in all cases. All the curves have been computed for the TE polarization and an oblique illumination at 20°.

# V. Experimental details

**Chemicals**

Oleylamine (OLA, Acros, 80-90%), %), lead chloride (PbCl$_2$, Afla Aesar, 99%), sulfur powder (S, Afla Aesar, 99.5%), oleic acid (OA, Afla Aesar, 90%), trioctylphosphine (TOP, Afla Aesar, 90%), n-hexane (VWR), Ethanol (VWR, >99.9%), Toluene (Carlo Erba, >99.8%).

**PbS nanocrystal synthesis**

In a three-neck flask, 300 mg of PbCl$_2$, together with 100 µL of TOP and 7.5 mL of OLA are degassed, first at room temperature and then at 110 °C for 30 min. Meanwhile, 30 mg of S powder is mixed with 7.5 mL of OLA until full dissolution and an orange clear solution is obtained. Then under nitrogen at 80 °C, this solution of S is quickly added to the flask. After 2 minutes, the reaction is quickly quenched by addition of 1 mL of OA and 9 mL of hexane. The nanocrystals are precipitated with ethanol and redispersed in 5 mL of toluene. This washing step is repeated one more time and the pellet is this time dispersed in 10 mL of toluene with a drop of OA. The solution is then centrifugated as is, to remove the unstable phase. The supernatant is precipitated with methanol and redispersed in toluene. Finally, the PbS NC solution in toluene is filtered through a 0.2 µm PTFE filter. The obtained solution is used for further characterization and device fabrication.

**Sample fabrication**

The array of Au square nanoparticles studied in Figure 3a-3e is defined by electron-beam lithography on a borosilicate glass substrate from Plan Optik coated by a bilayer of CSAR 62 positive resist and Electra 92 conductive coating from Allresist. The conductive coating is applied to avoid a charge accumulation during the writing procedure, the latter being performed with a PIONEER Two Raith system. After deposition of 2 nm of Ti and 25 nm of Au in a Plassys MEB 550S electron beam evaporator, followed by an overnight lift-off in butanone, the resulting Au nanoparticle array occupies an area of 200 × 200 μm². The scanning electron micrograph (SEM) of Figure S7 reveals that the actual size of the square nanoparticles is 415 nm × 415 nm instead of the nominal value of 400 nm × 400 nm. The measured period is 600 nm, as expected.

The fabrication of the photoconductive detector studied in Figure 3f starts with a first level of optical lithography (using an MJB4 mask aligner and the AZ 5214E photoresist), metal deposition (2 nm of Ti and 70 nm of Au) and lift-off to define the interdigitated electrodes. The spacing between two consecutive digits is 10 μm. These 10 μm wide channels are then filled with arrays of Au nanoparticles following the fabrication steps described above in this section. The total area of the photoconductive device is 870 × 1400 μm².

After defining the periodic arrays with and without interdigitated electrodes, the sample is introduced in a $N_2$-filled glovebox. A 15–20 nm thick layer of NCs is spun onto the sample at 2000 rpm, creating a uniform coating above the metallic patterns. Finally, the sample is dipped into a solution of EDT in ethanol for 60 s followed by a rinse in ethanol to strip the NCs from their native OA ligands and leave them instead cross-linked with shorter molecular chains.

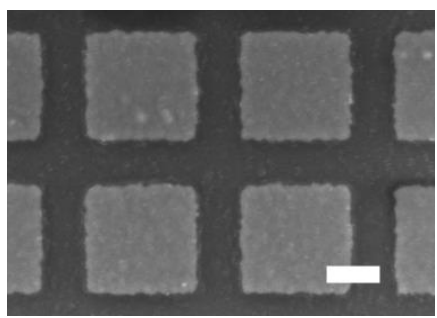

**Figure S7.** Scanning electron micrograph of the Au patterns. Scale bar 200 nm.

**Photoluminescence experiments**

All the PL experiments are performed with a customized BX51WI microscope from Olympus. Except for the results presented on Figure S11, we pump the sample with a continuous HeNe laser at 633 nm that is filtered with a BG40 filter from Thorlabs to remove the parasitic spontaneous emission in the near-infrared. This laser is focused to a micron-size spot using a 50X objective. For Figure S11, we have replaced this localized excitation by a wide field illumination scheme that excites a macroscopic part of the $200 \times 200$ µm$^2$ sample. This wide field pumping, covering a circular area with a diameter of 90 µm, is realized with the same objective and the wide field illuminator of our microscope. The light source attached to this illuminator is a tungsten lamp that is filtered to exclude the wavelengths outside the range 400 nm - 800 nm.

Our 50X objective (reference LCPLN50XIR) has a numerical aperture of 0.65 and it is also used to collect the infrared PL signal. A dichroic mirror (Thorlabs DMLP950R) is used to separate the IR signal above 950 nm from the visible pump. The remaining visible light that has not been filtered by the dichroic mirror is subsequently blocked with an RG780 longpass filter. The PL is analyzed with an Acton SP2356 imaging spectrograph coupled to a NIRvana InGaAs camera from Princeton Instruments (Figure S8).

- To record a PL spectrum, a 200 µm slit is placed at the spectrograph entrance. Light entering the spectrograph is then dispersed with a 85 groove/mm plane rule reflection grating. The raw spectra are then corrected by taking into account the transfer functions of the $50\times$ objective and the grating.

- To record a dispersion relation, the same configuration is used but an additional Bertrand lens is added in the optical path in a 4f configuration so as to image the back focal plane of the objective. The function of this lens is to Fourier-transform the signal, which is therefore analyzed in the reciprocal space.

- To record a back focal plane image, the reflection grating within the Acton SP2356 spectrograph is replaced by a flat silver mirror.

- For the polarization-resolved experiments, a wire grid polarizer (WP25M-UB from Thorlabs) is inserted in the optical path. We have carefully checked the effect of every optical elements on the polarization state of the PL signal, to make sure that no artifact affects the measurements.

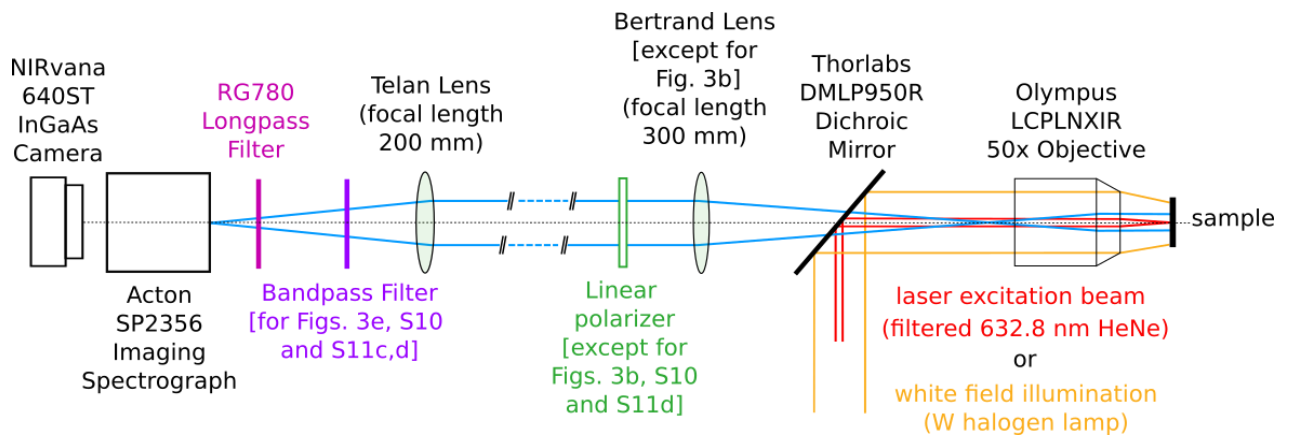

**Figure S8.** Schematic of the PL setup.

**Photocurrent measurements**

The interdigitated samples are biased at 7 V and illuminated with a 15X reverse Cassegrain objective using a water-cooled 150 W incandescent tungsten light source passing through a Fourier Transform Infrared (FTIR) spectrometer (Invenio-R from Bruker). The photocurrent that flows between the electrodes is amplified using a Femto DLCPA-200 amplifier, converted into a voltage and plugged back into the FTIR electronics. By normalizing the data with the spectrum measured with a calibrated 818-ST2 Ge photodiode from Newport, one obtains the responsivity of the sample.

# VI. Supplementary experiments

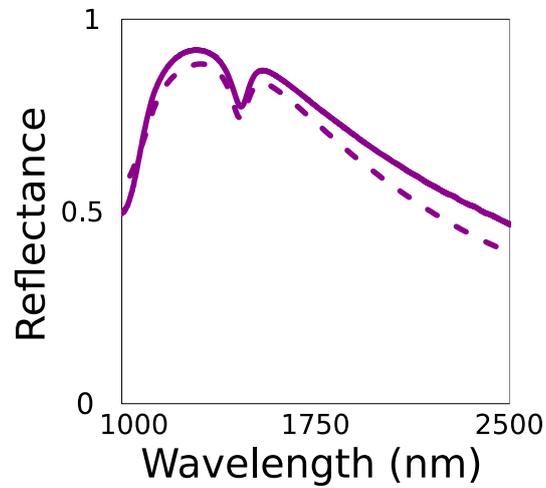

**Figure S9.** Experimental and simulated reflectance spectra of the sample studied in the main text, which consists in a periodic array of square Au nanoparticles coated with 2-3 layers of PbS NCs (continuous and dashed purple curves, respectively). This plot shows the same spectra as in Figure 3a, but over an extended range of wavelengths.

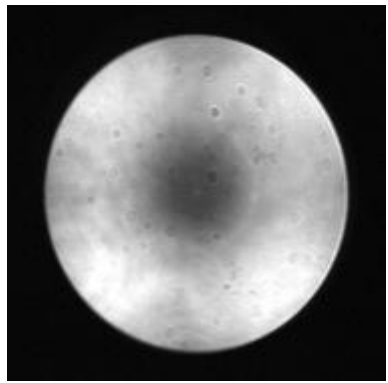

**Figure S10.** Back focal plane at 1450 nm without the linear polarizer inserted in the beam path. As in the main text, the experiment is realized by pumping the structure with a diffraction-limited 633 nm laser spot.

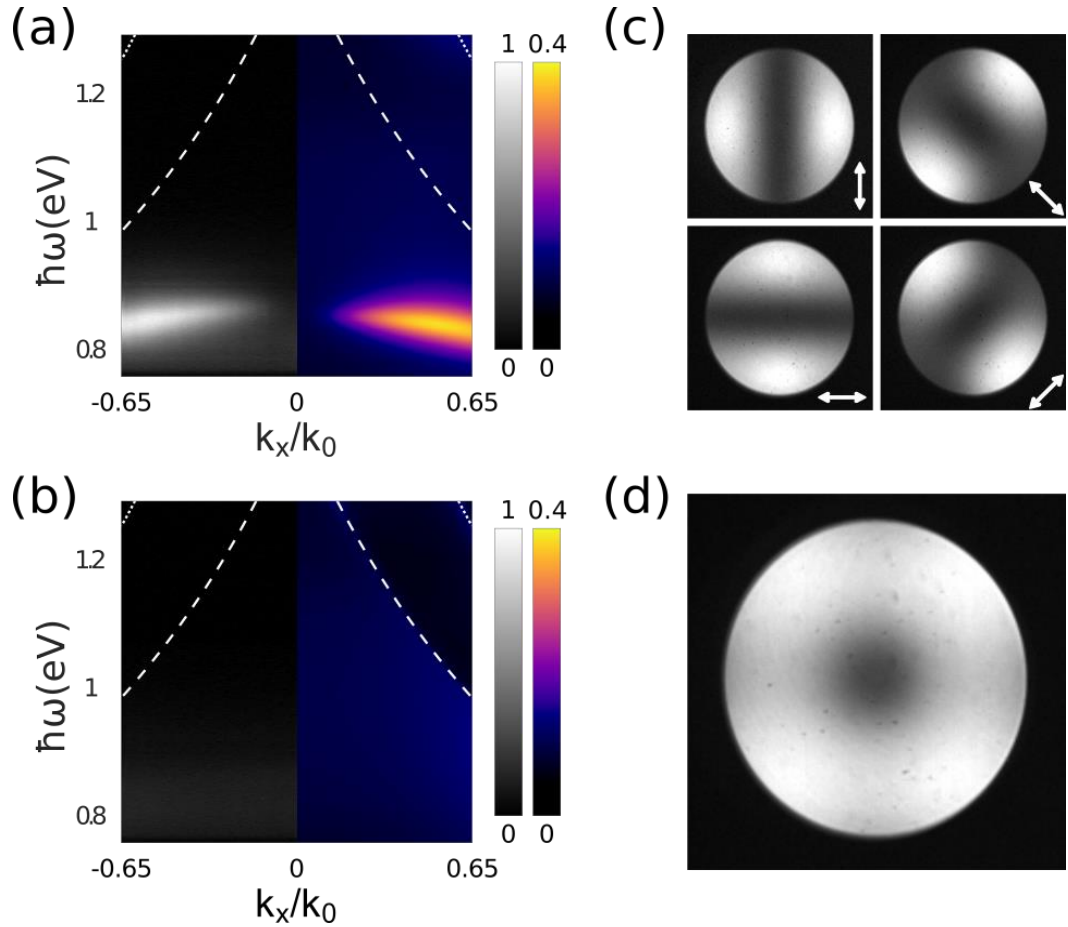

**Figure S11.** PL experiments realized with a wide field pumping. The sample is the same as the one studied on Figures 3 and S9-S10, except that the pumping is not realized with a diffraction-limited laser spot but by a white-light illumination scheme covering a macroscopic part of the sample (see page 10 of the present document for experimental details). Panels (a)-(b)-(c) plot the same quantities as those plotted on Figure 3c-3d-3e. Panel (d) plots the same quantity as in Figure S10. The results are essentially the same as those obtained with the diffraction-limited laser pumping. In fact, the wide field illumination creates an even nicer set of data because it averages the response of the array over a wide area.

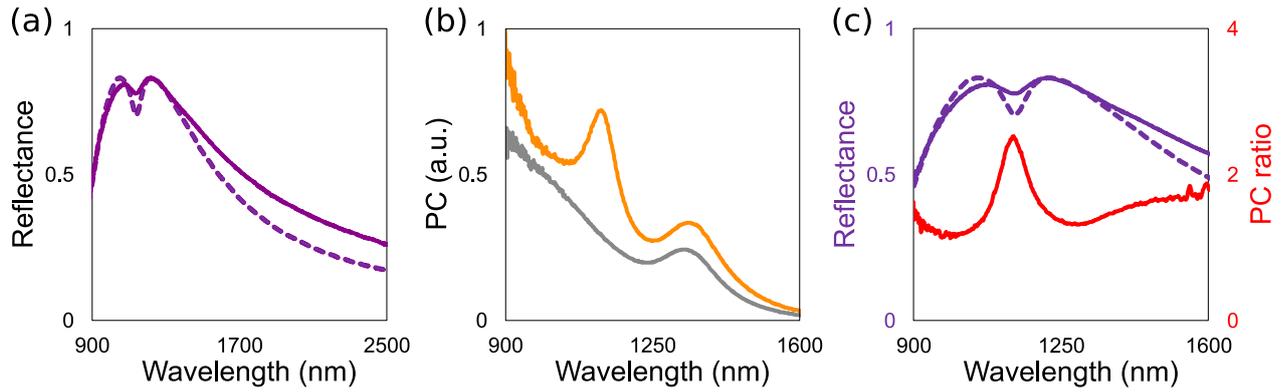

**Figure S12.** (a) Experimental and simulated reflectance spectra of a metasurface with a period P = 500 nm and square Au nanoparticles with a side of 300 nm, coated with a thin layer of PbS NCs (continuous and dashed purple curves, respectively). (b) Photocurrent spectra of PbS NC photoconductive detectors with (yellow curve) and without (gray curve) such a metasurface inserted between the interdigitated electrodes. The ratio between these two spectra is the red curve of panel (c), which also features a zoom of the curves already plotted on panel (a).

## VII. Supplementary references